# Optimal Design of Power Electronic Transformer based on Hybrid MMC under Boost-AC operation

Yaqian Zhang
School of Electrical Engineering
Southeast University
Nanjing China
yqzh@seu.edu.cn

Xudong Zhang
School of Electrical Engineering
Southeast University
Nanjing China
220232949@seu.edu.cn

Fujin Deng
School of Electrical Engineering
Southeast University
Nanjing China
fdeng@seu.edu.cn

Jianzhong Zhang
School of Electrical Engineering
Southeast University
Nanjing China
jiz@seu.edu.cn

*Abstract*—The bridge arm of the hybrid modular multilevel converter (MMC) is composed of half-bridge and full-bridge sub-modules cascaded together. Compared with the half-bridge MMC, it can operate in the boost-AC mode, where the modulation index can be higher than 1, and the DC voltage and the AC voltage level are no longer mutually constrained; compared with the full-bridge MMC, it has lower switching device costs and losses. When the hybrid MMC boost-AC mode is used in the power electronic transformer, the degree of freedom in system design is improved, and the cost and volume of the power electronic transformer system can be further reduced. This paper analyzes how to make full use of the newly added modulation index of freedom introduced by the boost-AC hybrid MMC to optimize the power electronic transformer system, and finally gives the optimal modulation index selection scheme of the hybrid MMC for different optimization objectives.

## I. INTRODUCTION

With the deepening of power electronics and the commercial maturity of various power electronic devices manufacturing technologies, power electronic transformers play a crucial role in modern power systems, such as new energy generation grid connection, AC and DC distribution grids, and electrified energy consumption side [1],[2]. Compared with power frequency transformers, power electronic transformers have the advantages of high power density, high power quality, power control function, interactive power transmission and fault ride-through. Through the introduction of high-frequency transformer to replace the traditional power frequency transformer, as well as a variety of power electronic converter combination and cascade, the power electronic transformer can not only achieve the traditional industrial frequency transformer voltage rise and fall and power transmission function, but also can achieve a variety of flexible system-level control, such as current control, to improve the intelligence, flexibility, and controllability of the grid-connected system [3], [4].

The high-voltage side of the power electronic transformer is directly connected to the medium- and high-voltage grid, and presently, the grid-connected side generally adopts the cascaded bridge multilevel converter or modular multilevel converter (MMC) [5-7]. In contrast, MMC-based power electronic transformers with a common medium or high voltage DC bus avoids the constraint of the same number of DC/DC units and the sub-modules, resulting in a larger optimisation space and higher flexibility. Fig. 1 shows a typical MMC-based power electronic transformer structure, in which MMC power stage and DC/DC power stage only through the medium voltage DC bus to achieve active power coupling and voltage coupling.

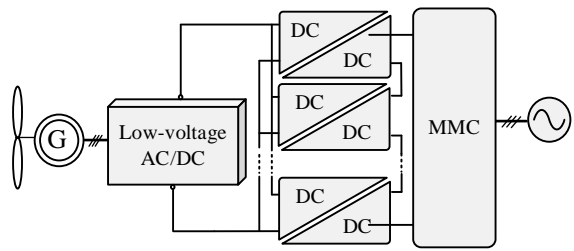

Fig. 1. Grid-connected power electronic transformer system for wind turbines based on MMC.

Currently, among the three typical MMC topologies in Fig. 2, the half-bridge sub-module(HBSM) is the most widely applied MMC sub-module structure, which has the advantages of low cost and low loss [8-10]. However, it can only work under $U_{dc}/2 \geq U_{ac}$ [3].

Due to the demanding of high AC power quality, the modulation index of HB-MMC cannot exceed 1, which means the half of the middle-voltage DC bus voltage cannot be lower than the AC phase voltage amplitude. Therefore, when the HB-MMC is applied in power electronic transformers, the selection of the middle-voltage DC bus voltage is constrained by the AC voltage, which in turn limits the design of the DC/DC power stage, including the number of DC/DC units, the voltage level, and the power level. Taking a 35 kV AC distribution grid (AC phase voltage amplitude of 28 kV) as an example, the middle-voltage DC bus voltage of the HB-MMC cannot be lower than 56 kV, which causes a big challenge to the internal insulation of the power electronic transformer and brings the large number of DC/DC units and the high complexity and cost of the system.

In contrast, when full-bridge sub-module (FBSM) are applied in MMC, such as hybrid MMC and FB-MMC inFig. 2, for the reason that the FBSM are capable of outputting a negative level, and the bridge arm voltages are able to be expanded to bipolar, FB-MMC and hybrid MMC are able to realise an operation mode with a modulation index more than 1, which is called the boost-AC operation[11], as shown in Fig. 3. In the boost-AC mode, the AC phase voltage amplitude of hybrid MMC is higher than the half of the DC bus voltage, and the output voltage maintains a linear relationship with the input voltage, and the AC voltage and current quality will not be influenced [12]. The boost-AC operation capability breaks through the mutual restriction between the DC bus voltage and AC voltage of the MMC, which enables the DC bus voltage of the power electronic transformer to vary over a wide range but does not affect the AC voltage output. Therefore, the application of FB-MMC and hybrid MMC into power electronic transformer systems can increase the DC bus voltage as a new design free degree and improve the optimisation potential of power electronic transformers.

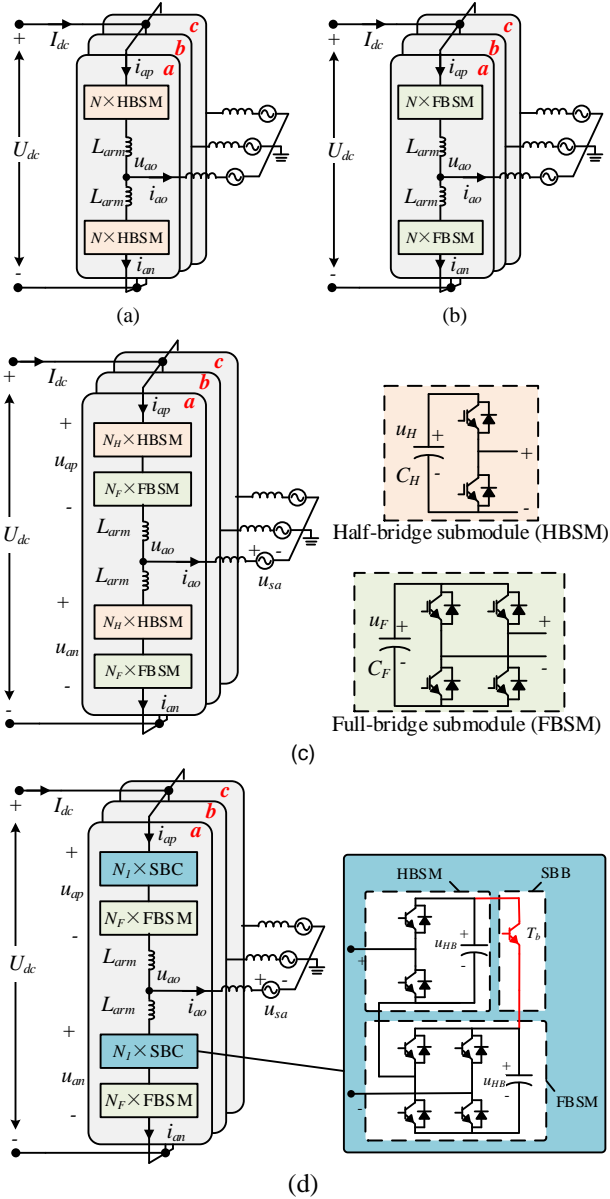

Fig. 2. Three typical MMC topologies. (a) HB-MMC. (b) FB-MMC. (c) Traditional Hybrid MMC. (d) Hybrid MMC with self-balancing branch(SBB).

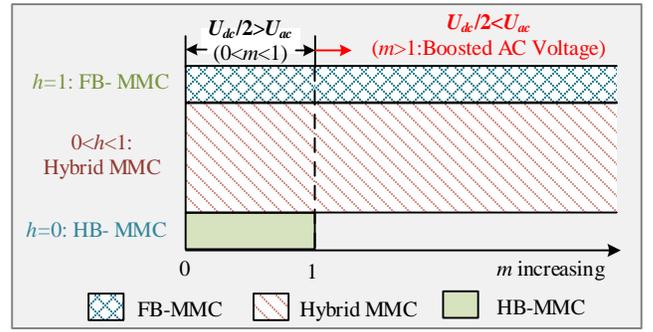

*$m$: modulation index of MMC ($2U_{ac}/U_{dc}$); $h$: hybridization ratio of FBSMs.

Fig. 3. Modulation index range of different MMCs. (0<$m$<1: hybrid MMC, HB-MMC, and FB-MMC; $m$>1: FB- and hybrid MMC.)

In paper[13], Camurca et al. discussed the impact of the over-modulated operation capability of FB-MMC on the performance of power electronic transformers, and the analyses showed that the introduction of over-modulated operation can effectively reduce the cost of power electronic transformers, but most of the comparisons were qualitative analyses, and the optimal design method for the medium-voltage dc bus was not given. In the paper[14], Nakanishi et al. operate the FB-MMC at a very high modulation index, so that the middle-voltage DC bus voltage level is sufficiently reduced, and the voltage ratio of the DC/DC power stage can be as low as 1:1, and at this time, the volume of the power electronic transformer can be reduced to 90% compared with the traditional grid-connected system with an power frequency transformer, but the paper only considers one kind of design of the modulation index of FB-MMC and does not compare the effects of different modulation index and different voltage of middle-voltage MVDC bus on the total volume of the power electronic transformer. In conclusion, all the current applications for over-modulated operation only discuss the FB-MMC. Compared with the FB-MMC, the hybrid MMC can also achieve over-modulated operation, but it has fewer number of FBSMs, lower cost, and lower operating losses, so applied of the hybrid MMC in the power electronic transformer will gain more advantages in terms of cost and losses.

Based on the design free degree of the middle-voltage DC bus of power electronic transformers, combined with the design of DC/DC power stages at different voltage of middle-voltage DC bus, this paper analyses the impact of different voltage of middle-voltage DC bus on the performance of the power electronic transformer system in detail, including the system cost, volume, and loss, so as to propose the optimal middle-voltage DC bus design scheme. Finally, the future research perspectives are presented.

## II. POWER ELECTRONIC TRANSFORMER BASED ON BOOST-AC HYBRID MMC

### A. System Structure and Parameters

The hybrid MMC power electronic transformer based wind turbine grid-connected structure is shown in Fig. 4, including the detailed structure of the DC/DC unit and the hybrid MMC, and its related parameters are shown in Table I. The AC output voltage on the wind turbine side is 3 kV and the rectified low voltage DC bus is 5 kV, which is used as the input to the DC/DC power stage. The rated power of the power electronic transformer in this paper is set to 5 MW, which is consistent with the current rated power of the 3 kV wind turbine, and then the rated power of both the hybrid MMC and the DC/DC power stage are 5 MW.

The DC/DC power stage consists of DC/DC converter units with parallel inputs and series outputs, which can achieve voltage boosting and low and medium voltage electrical isolation. The low-voltage DC bus voltage of the DC/DC unit is 5 kV, and the existing commercial power electronics switching devices are not yet able to meet the 5 kV bus voltage through the two-level voltage converter, so the DC/DC unit selects the diode-clamped dual active bridge converter structure based on diode clamp[15], the detailed topology, the corresponding modulation strategy and control methods can be referred to the paper[16]. The parameters of the power electronic transformer and each power stage are shown in Table 1, and $N_{DC}$ is the number of DC/DC units in cascade, and the operating voltage of each switch of the DC/DC unit is 2.5 kV, and IGBTs with the voltage of 4.5 kV are selected as the switching devices of the DC/DC unit with a switching frequency of 2.5 kHz in this chapter.

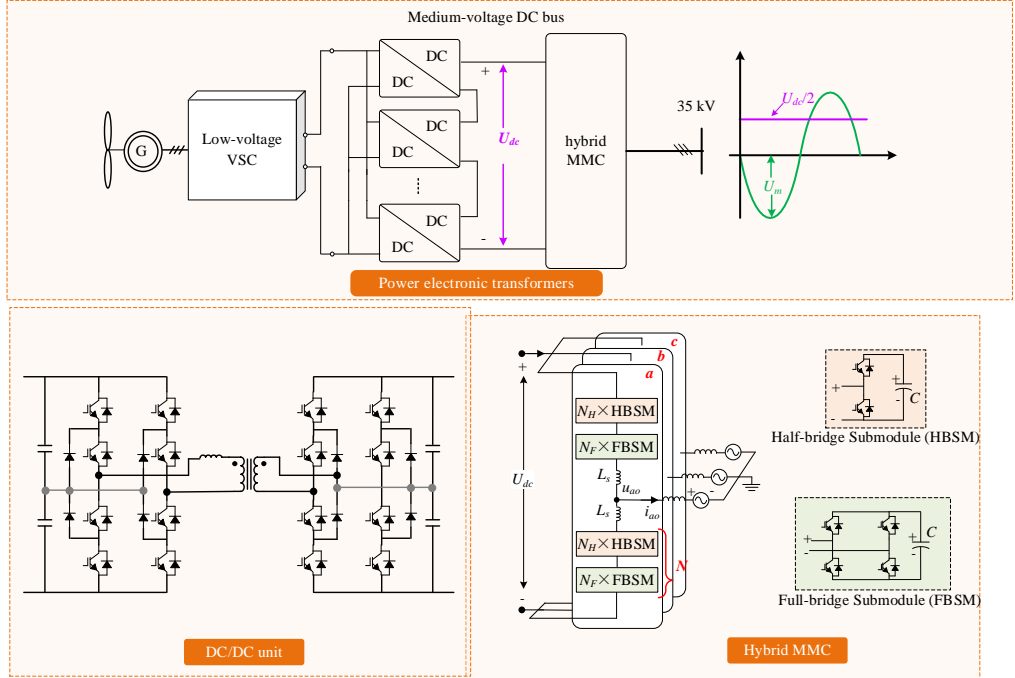

Fig.4 Grid-connected power electronic transformer for wind power generation based on hybrid MMC.

TABLE I. POWER ELECTRONIC TRANSFORMERS AND PARAMETERS OF EACH POWER STAGE

| Power electronic transformers | | DC/DC Units | | Hybrid MMC | |
|---|---|---|---|---|---|
| *Parameter* | *value* | *Parameter* | *value* | *Parameter* | *value* |
| Rated power | 5 MW | Input voltage of single turbine | 5 kV | Voltage of sub-module capacitor | 2 kV |
| Output voltage of turbine | 3 kV | Output voltage of single turbine | 5 kV | Voltage of middle-voltage AC | 35 kV |
| Voltage of low-voltage AC | 5 kV | Ratio of transformer | 1:1 | Rated power | 5 MW |
| Voltage of middle-voltage AC | 35 kV | Frequency of transformer | 5 kHz | | |
| | | Rated power | 5 MW/$N_{DC}$ | | |

TABLE II. IGBT SELECTION FOR HYBRID MMCS WITH DIFFERENT MODULATION RANGES (INFINEON)

| Range of modulation index | Voltage of medium-voltage DC bus | IGBT model | Voltage stress | Current stress |
|---|---|---|---|---|
| 1<$m$<18 | 3.3 kV <$U_{dc}$< 60 kV | FZ800R45KL3_B5 | 4.5 kV | 800 A |
| 18<$m$<23 | 2.6 kV <$U_{dc}$< 3.3 kV | FZ1000R45KL3_B5 | 4.5 kV | 1000 A |
| 23<$m$<28 | 2.1 kV <$U_{dc}$< 2.6 kV | FZ1200R45KL3_B5 | 4.5 kV | 1200 A |

TABLE III. IGBT SELECTION FOR DC/DC UNITS WITH DIFFERENT MODULATION RANGES (INFINEON)

| Range of modulation index | Voltage of medium-voltage DC bus | IGBT model | Voltage stress | Current stress |
|---|---|---|---|---|
| 1<$m$< 6 | 10 kV< $U_{dc}$ < 60 kV | FZ800R45KL3_B5 | 4.5 kV | 800 A |
| 6<$m$<7.5 | 8 kV< $U_{dc}$ < 10 kV | FZ1000R45KL3_B5 | 4.5 kV | 1000 A |
| 7.5<$m$<9 | 6.67 kV< $U_{dc}$ < 8 kV | FZ1200R45KL3_B5 | 4.5 kV | 1200 A |
| 9<$m$<11.2 | 5.35 kV< $U_{dc}$ < 6.67 kV | FZ1500R45KL3_B5 | 4.5 kV | 1500 A |

TABLE IV. COMPARISON OF THREE TYPES OF MMCS

|  | Traditional hybrid MMC (1<$m$<2) | Hybrid MMC with SBB (1<$m$<7) | FBSM MMC ($m$>1) |
|---|---|---|---|
| Modulation index range | ▲▲ | ▲▲▲ | ▲▲▲▲ |
| Power Density | ▲▲▲▲ | ▲▲ | ▲▲▲ |
| Efficiency | ▲▲▲ | ▲▲▲▲ | ▲▲ |
| Lower cost | ▲▲▲▲ | ▲▲▲ | ▲▲ |

The medium and high voltage DC bus is connected in parallel with the DC side of the hybrid MMC, and the AC side of the hybrid MMC is connected to the 35 kV grid, so the AC voltage level is unchanged. For optimal design of the power electronic transformer, this paper focuses on the medium-voltage DC bus. Since the parameters of the hybrid MMC, including the sub-module capacitance $C$ and the number of sub-modules $N$ (where the number of half-bridge sub-modules and full-bridge sub-modules are $N_H$ and $N_F$, which satisfy the relationship $N_H+N_F=N$) are closely related to the DC bus voltage, and the specific relationship can be referred to [17]. Table I only gives the parameters of the hybrid MMC which are not affected by the DC bus voltage or modulation index, e.g., the rated voltage of sub-module capacitor is 2 kV.

### B. Device Selection Based on Boost AC Hybrid MMC

The selection of Insulate-Gate Bipolar Transistor (IGBT) for the hybrid MMC and DC/DC units at different medium voltage DC bus voltages, i.e., different hybrid MMC modulation index, are shown in Tables II and III, respectively.

It can be found that if no over-modulation operation is introduced, the middle-voltage DC bus voltage is 60 kV, and without redundancy, a cascade of at least 12 DC/DC units ($N_{DC}$=12) is required to obtain a middle-voltage DC bus voltage. On the other hand, under a specific power level, the middle-voltage DC bus voltage of the power electronic transformer is high and the current is low, while the capacity of existing high-voltage commercial switching devices far exceeds the capacity of required switching devices. For example, the medium-voltage DC bus voltage is $U_{dc}$ = 60 kV, the current of switching devices of DC/ DC units is only 83 A, and 4.5 kV IGBT of Infineon minimum current stress is 800 A, which means the IGBT selection and the actual current stress does not match well, resulting in a serious waste of devices. The introduction of boost-AC can improve the design flexibility of the middle-voltage DC bus voltage and add a new free degree for the optimal design of power electronic transformers.

### III. COMPREHENSIVE SYSTEM ASSESSMENT

In this paper, three kinds of MMC with boost-AC capability are used as examples, including the traditional hybrid MMC [18], the balanced branch-based hybrid MMC [19] and the FB-MMC [20], to analyse the total volume and cost of the power electronic transformer over a wide range of modulation index. It should be noticed that the modulation index of traditional hybrid MMC is limited by the capacitor voltage balance with a range of 1<$m$<2, while the modulation index of improved balanced branch MMC and full-bridge MMC has a range of m>1.

### A. Basic principles of comparative analysis

The principles for calculating the cost and volume of the power electronic transformer system are as follows. Since the turbine-side rectifier is not affected by the regulation system, only the DC/DC power stage and the MMC power stage are considered here, and the turbine-side rectifier power stage is not considered. Firstly, the MMC volume and cost include the sub-module capacitor part and the IGBT part, the cost and volume of the sub-module capacitor part are proportional to the total capacitance value, and the capacitance design of the three kinds of MMC can be referred to the corresponding paper, while the cost of the IGBT part is proportional to the volume and the number of the IGBT. Secondly, the volume and cost of the DC/DC power stage include the high-frequency transformer part and the IGBT part, where the volume and cost of the IGBT part are proportional to the number of IGBT, and the cost of the high-frequency transformer part is not affected by the modulation index and the volume is proportional to the number.

### B. Comparison of volume and cost

Based on the above principles, Fig. 5 and Fig. 6 show the trend of total volume and total cost of power electronic transformers for different modulation index (different middle-voltage DC bus voltages), respectively.

(1) $U_{DC}$= 40 kV, $m$=1.5

The volume of the power electronic transformer system is reduced by 76% and the cost is reduced by 80% when the modulation index m=1.5 with a middle-voltage DC bus voltage of 40 kV.

(2) $U_{DC}$ = 30 kV, $m$=2

If the middle-voltage DC bus voltage is reduced to 30 kV, the number of DC/DC units is reduced to 1/2, and the volume and cost are reduced to 55% and 65%, respectively; the volume of the power electronic transformer is reduced to 75%, and the cost is reduced to 75%.

(3) $U_{DC}$= 20 kV, $m$=3

The middle-voltage DC bus voltage is reduced to 20 kV, the number of DC/DC units is reduced to 1/3, and the cost and volume of the DC/DC power stage are reduced to 53% and 40%, respectively; since the volume of the hybrid MMC increases by 27% at this point, the final volume of the power electronic transformer is increased by a factor of 1.05, and the cost is reduced by 69%.

(4) $U_{DC}$= 15 kV, $m$=4

If the middle-voltage DC bus voltage is further reduced to 15 kV, the number of DC/DC units is reduced to 1/4, and the cost and volume are reduced to 48% and 30%, respectively; the volume of the hybrid MMC power stage is increased by a factor of 1.64, which is about 98% of the volume; and the volume of the power electronic transformer is increased by a factor of 1.35, and the cost is reduced by 68%.

(5) $U_{DC}$= 10 kV, $m$=6

Further, when the middle-voltage DC bus voltage is reduced to 10 kV, the number of DC/DC units can be as low as 1/6, while the cost and volume of the DC/DC power stage are reduced to 43% and 25%; at this time, the volume of the MMC increases by a factor of 2.3, and the cost increases by a factor of 1.17. The cost of the power electronic transformer is reduced to 71%, while the volume increases to 1.83 times.

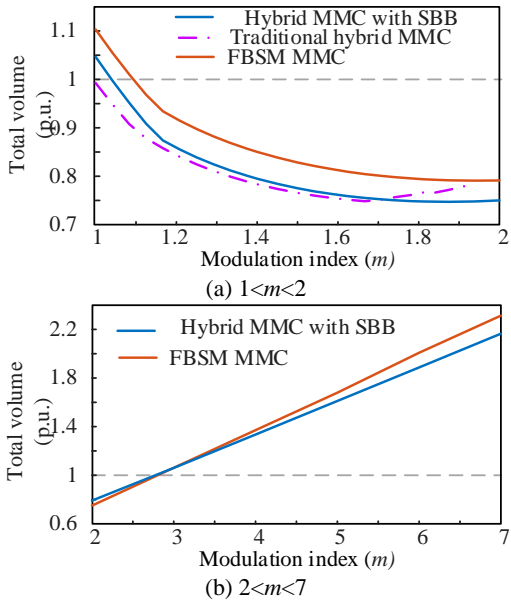

Fig.5 Total volume of power electronic transformers.

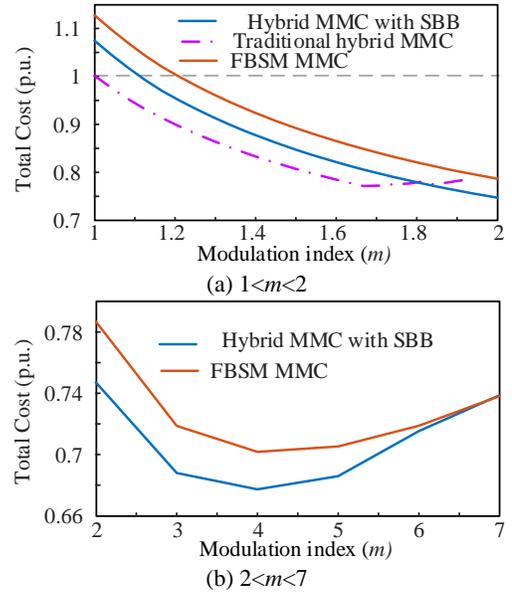

Fig.6 Total cost of power electronic transformers.

### C. Comparison of power losses

As shown in Fig. 7, the two types of hybrid MMCs has much lower power losses than the FBSM MMC. Especially, the hybrid MMC can achieve the same wide modulation range but lower power losses compared with the FBSM MMC. It should be noticed that the power losses of all the boost-AC-mode MMC are increased with the increase of modulation index because the arm currents are increased by the increased DC current under the same power level.

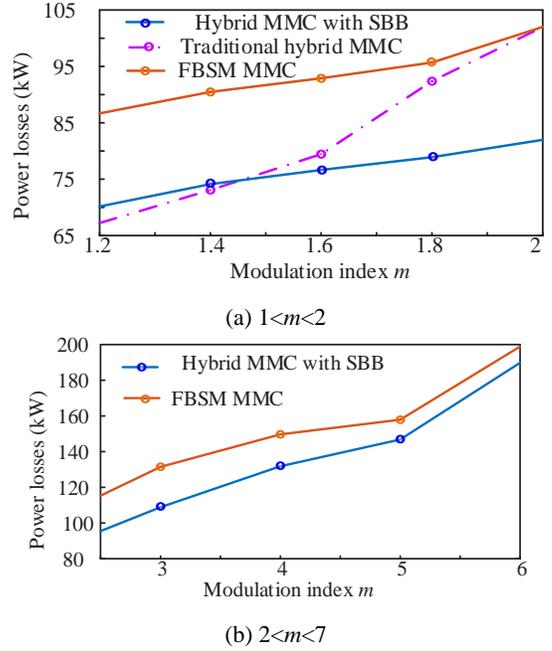

Fig.7 Power losses of MMCs.

### IV. DISCUSSION AND SUMMARY

The comparisons of different MMCs with the capability of boost AC mode are listed in Table IV. It is shown that the FBSM

MMC has largest range of modulation indexes, the hybrid MMC with Self-balancing Branch (SBB) performs best in terms of the efficiency, and the traditional hybrid MMC has lowest cost and highest power density within the modulation index range of $1<m<2$.

For different power electronic transformer optimisation objectives, the middle-voltage DC bus voltage is designed differently:

(a) When considering volume reduction as the objective, the modulation index of hybrid MMC is optimally set at $1.6<m<1.8$, i.e., the DC bus voltage should be set at 31 kV to 35 kV, at which time there is no clear difference in the volume of the MMC power stage between applying the traditional hybrid MMC topology and the hybrid MMC topology based on the improved balanced branch, and the utilization of boost-AC mode can reduce the total volume of the power electronic transformer to 75%.

(b) When considering cost reduction as the objective, the optimal range of modulation index for hybrid MMC is $3 < m < 5$. In this case, the MMC power stage applying hybrid MMC based on improved balanced branching reduces the total cost up to 68% compared to the power electronic transformer without the introduction of over-modulation, while the volume may increase up to 1.4 times.

However, the over-modulation of the hybrid MMC is not suitable for the efficiency optimisation of the power electronic transformer. The loss in the MMC power stage will increase with the increase of DC bus current when the modulation index increases, which is not conducive to efficiency improvement.